\documentclass[twocolumn,american,aps,pra]{revtex4-1}
\usepackage[T1]{fontenc}
\usepackage[utf8]{luainputenc}
\usepackage{color}
\usepackage{float}
\usepackage{amsmath}
\usepackage{graphicx}
\usepackage{babel}
\begin{document}

\title{Theoretical investigation of in situ k-restore processes for damaged ultra-low-k dielectrics}

\author{Anja F{\"o}rster$^{*,\#}$, Christian Wagner$^{\#}$, J{\"o}rg Schuster$^{*}$,
Sibylle Gemming$^{\#,+}$}

\affiliation{{*} Fraunhofer ENAS, Technologie-Campus 3, 09126 Chemnitz, Germany}

\affiliation{\# TU Chemnitz, Reichenhainer Str. 70, 09126 Chemnitz, Germany}

\affiliation{+ Helmholtz-Zentrum Dresden - Rossendorf, Institute of Ion Beam Physics
and Materials Research, Bautzner Landstraße 400, 01328 Dresden, Germany}
\begin{abstract}
Ultra-low-k (ULK) materials are essential for today's production of
integrated circuits (ICs). However, during the manufacturing process
the ULK's low dielectric constant (k-value) increases due to the replacement
of hydrophobic species with hydrophilic groups. We investigate the
use of plasma enhanced fragmented silylation precursors to repair
this damage. The fragmentation of the silylation precursors \foreignlanguage{american}{octamethylcyclotetrasiloxane}
(OMCTS) and \foreignlanguage{american}{bis(dimethylamino)-dimethylsilane}
(DMADMS) and their possible repair reactions are studied using density
functional theory (DFT) and molecular dynamics (MD) simulations. 
\end{abstract}
\maketitle

\section{Introduction}

Gordon Moore predicted that the number of transistors per integrated
circuit (IC) would double biennially \cite{01}. To achieve
this goal today's microelectronic industry uses materials with low
dielectric constants (k-value): ultra-low-k (ULK) materials. Pores
and hydrophobic species in the ULK materials are the reason for their
low k-value. However, the hydrophobic species, in particular methyl
groups, are prone to be replaced by hydrophilic, polar groups during
the manufacturing process due to active radicals and highly energetic
vacuum-ultra-violet photons \cite{02}. The substitution
of $\textrm{Si-CH}{}_{3}$ bonds with Si\nobreakdash-H bonds (H-damage)
leads to an increased k-value. In air contact, this H-damage can further
react to an OH\nobreakdash-damage by forming Si-OH bonds \cite{03}
or lead to adsorbed water molecules \cite{04}.

Three repair mechanisms to re-decrease the k-value were suggested
in literature \cite{02}: The UV assisted thermal curing,
the silylation process and the k-restore via hydrocarbon plasma. Either
of the these mechanisms has its own assets and drawbacks. The UV assisted
thermal curing removes the hydroxyl groups and water molecules from
the material by breaking the bonds with UV radiation and high temperatures.
However, in order to cure the damage completely, temperatures of about
$600-1000\deg$ C are required \cite{02}. Because of the ULK's
porous structure, this results in the compression or collapse of the
material \cite{05}. 

The k-restore via hydrocarbon plasma is based on methane fragments.
These fragments are supposed to diffuse into the porous material to
repair the damage by replacing undesired Si-OH bonds with $\textrm{Si-CH}{}_{3}$
bonds. Experiments showed that the fragments instead build a carbon
rich layer on the surface \cite{06}, thus no repair takes place.
However, the carbon rich layer can protect the ULK material from further
damage.

The silylation precursors, on the other hand, replace the damaged
bonds with $\textrm{Si-CH}{}_{3}$ bonds at low temperatures ($T<300 \deg$C).
However, the size of the precursors (6-9 \AA{}) prevents them from diffusing
into deeper regions of the damaged ULK material. Therefore, the repair
via silylation is limited to the surface \cite{07}. 

An obvious alternative for the in-situ k-restore is the plasma fragmentation
of silylation precursors, the so-called plasma repair process. This
repair process is distinguished as it combines the good repair behavior
of the silylation process with the good diffusion be\textcolor{black}{havior
of the hydrocarbon plasma, while at the same time only requiering
a temperature of $80\deg$C \cite{08}.}

\textcolor{black}{In our theoretical study we analyze the reaction
energies of numerous fragmentations and selected repair reactions.
Based on these findings, fragmentation and reaction pathways can be
identified, which are likely to occur in experiments. Finally we compare
our results with the experimental investigation of K{\"o}hler et al. \cite{08}. }

\begin{figure*}
\includegraphics[width=1.\textwidth]{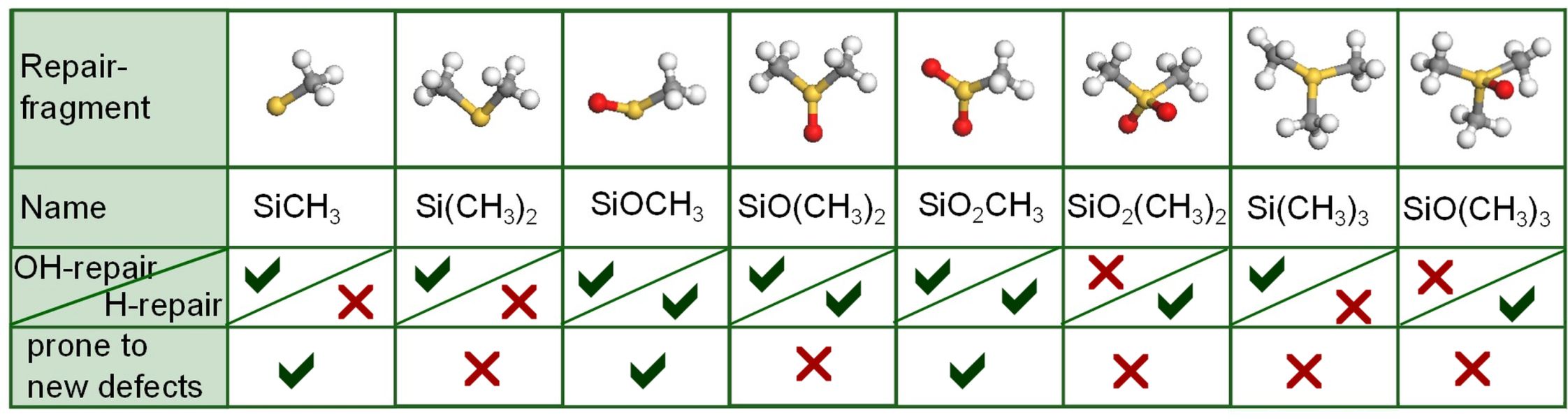}
\protect\caption{Overview of the studied repair fragments. Their name, the damage they
can cure and their susceptibility to create new defects are listed.\label{fig:Overview-Repair-Fragments}}
\end{figure*}

\section{Repair Fragments}

The combination of the advantages of both the silylation and the hydrocarbon
plasma process in the plasma repair process is achieved by the plasma
enhanced fragmentation of silylation precursors. For our theoretical
analysis we considered eight different fragments which are shown in
figure \ref{fig:Overview-Repair-Fragments}. They can be obtained
from the silylation precursors \foreignlanguage{american}{bis(dimethylamino)-dimethylsilane
}(DMADMS, figure \ref{fig:Fragmentation-of-DMADMS}a) and \foreignlanguage{american}{octamethylcyclotetrasiloxane}
(OMCTS, figure \ref{fig:Fragmentation-of-DMADMS}b).

As can be seen in figure \ref{fig:Exemplary-silylation-and}a, the
silylation and plasma repair processes display an identical repair
behavior. This includes the disadvantage that some silylation precursors/plasma
repair fragments can only repair one damaged site, resulting in a
steric hindrance for other repair processes in the close vicinity.
However, the fragments from the plasma repair process are about half
the size of the silylation precursors (2.5-5 Å vs. 6-9 Å) and thus
can repair the deeper region of the damaged ULK materials. Further,
the plasma repair process does not result in additional molecules
that are a possible source for further damage if they do not evaporate
from the ULK material. 

Also, opposed to the silylation process, the plasma repair process
can cure H\nobreakdash-damages in addition to OH\nobreakdash-damages.
It can be deduced from figure \ref{fig:Overview-Repair-Fragments}
that not all repair fragments can repair both damage types. This is
due to the H\nobreakdash-repair depending on oxygen atoms, whereas
the OH\nobreakdash-repair requires the silicon atom to have at least
one dangling bond. Further, if the repair fragment contains exactly
one methyl group, then the fragment is prone to further damage.

\section{Computational Details}

We used both density functional theory (DFT) and molecular dynamics
(MD) to investigate the fragmentations of the silylation precursors
and the repair reactions of the gained repair fragments. For the DFT
calculations we used Dmol$^3$ \cite{09,10} as implemented
in Materials Studio (Accelrys, Version 6.0) \cite{11}. We chose
the PBE functional in combination with the DNP(3.5) basis
set \cite{12}. The convergence criteria were set to $10^{-5}$~Ha
for the geometry optimizations and to $10^{-6}$~Ha for the scf cycles.
In cases of problematic convergence thermal smearing up to 0.008 Ha
was activated. Because the automatic orbital cutoff value of methyl
being below the value for silicon containing species, it was fixed
to 4.6 Å for all calculations. The Grimme DFT-D correction was also
used. 

For the MD calculations we used Gulp \cite{13} as implemented in
Materials Studio (Accelrys, Version 6.0) with the ReaxFF 6.0 force
field \cite{14,15,16,17},
as well as Lammps \cite{18} with the ReaxFF force field as parametrized
by Kulkarni et. al \cite{19}. With the former program
geometry structures were optimized until a stationary point was reached.
Using Lammps a dynamical calculation at a temperature close to 0 K
was performed until a convergence of $10^{-6}$~Ha was reached, followed
by a minimization at 0K with a convergence criteria of $10^{-10}$~Ha.

We compare the DFT and MD results to determine the reliability of
the used force fields. This is done with the aim to study larger clusters
and surfaces in future studies as such systems are too large to be
handled with DFT methods. If not otherwise stated, all reaction energies
were calculated for temperatures of 0 K.

\section{Model System}

\begin{figure}[b]
\includegraphics[width=0.47\textwidth]{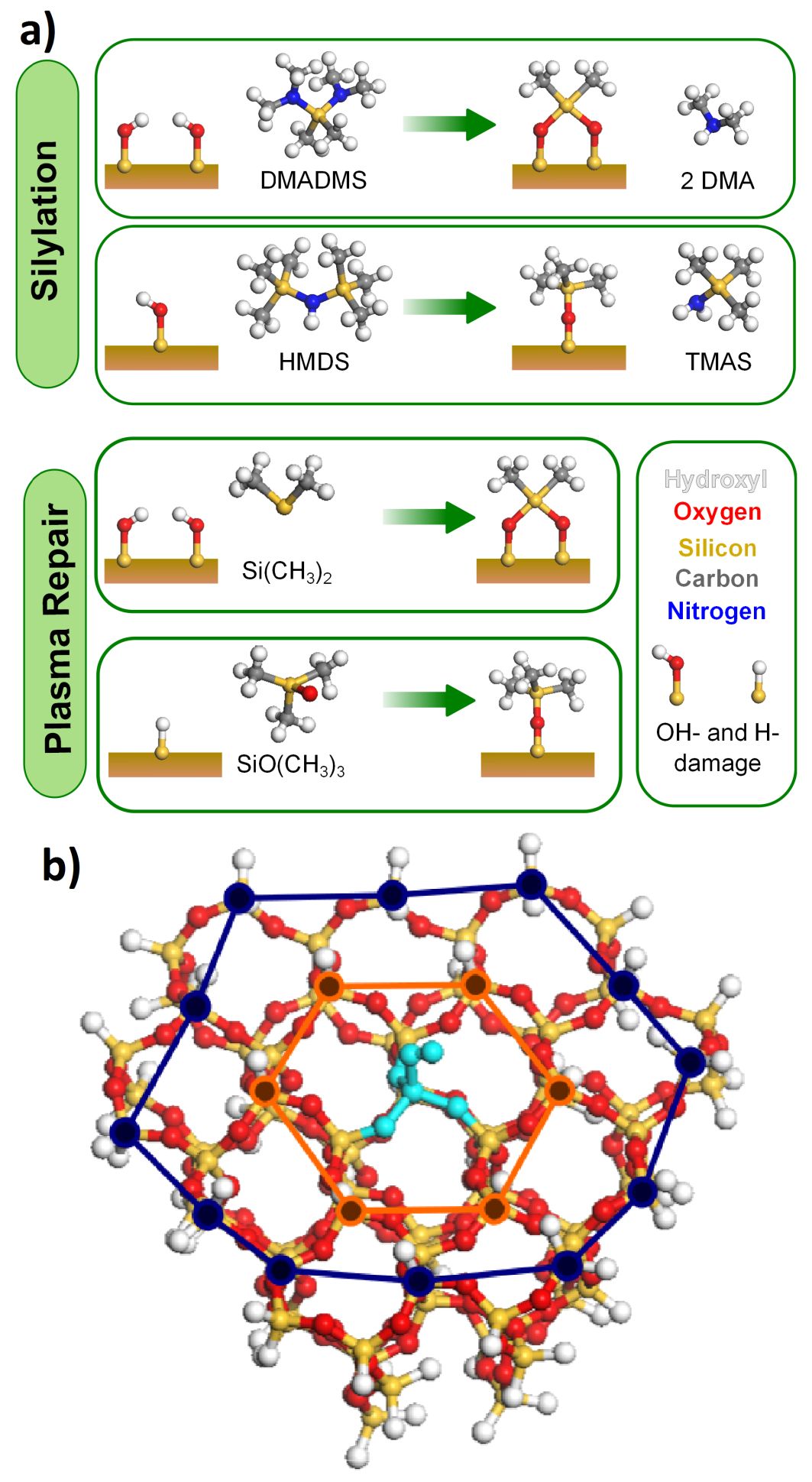}

\protect\caption{a) Exemplary silylation and plasma repair processes.
The silylation process with DMADMS and HMDS (
also results in residual molecules dimethylamine
(DMA) and trimethylaminosilane (TMAS) in comparison
to the plasma repair process with $\textrm{Si(C}\textrm{H}_{3}\textrm{)}_{2}$
and $\textrm{SiO(C}\textrm{H}_{3}\textrm{)}_{3}$.\\
b) the silicon oxide cluster model for the OH-damage
The atoms highlighted in light blue show the OH-damage region. The
hexagons on the cluster illustrate the freely optimized silica dodecagons
that provide a steric hindrance (orange) and the geometrically constrained
boundary silica dodecagons which stabilize the structure (dark blue).}
\label{fig:Exemplary-silylation-and}
\end{figure}

In order to mimic the high energy of the plasma-generated
fragments of the silylation precursors, we study the fragmentation
reactions at a temperature of $700\deg$ C. We limit the studied fragmentation
reactions to the ones that result in our desired plasma repair fragments
from figure \ref{fig:Overview-Repair-Fragments}. Further, we restrict
the reactions to those leading to exactly one or four repair fragments
with possible closed-shell residues for the DMADMS and the OMCTS fragmentation,
respectively.

We constructed a silicon oxide cluster model to investigate the effectiveness
of selected repair fragments, the influence of surrounding atoms on
the repair behavior and the reliability of the used force fields in
comparison to DFT results. Thus, the silicon oxide cluster consists
of the damaged site and two rings of surrounding silica atoms which
mimic steric hindrance and stabilize the structure (see figure \ref{fig:Exemplary-silylation-and}b).

With a sum formula of $\textrm{S}\textrm{i}_{86}\textrm{O}_{135}\textrm{H}_{73}\textrm{-}\textrm{OH}$
and $\textrm{S}\textrm{i}_{86}\textrm{O}_{135}\textrm{H}_{73}\textrm{-}\textrm{H}$
for the OH- and H-damage, respectively, our cluster is at the limit
of a computationally reasonable size for DFT. Therefore, to decrease
the computational costs for the DFT calculations the 73 hydrogen atoms
of the cluster (see figure \ref{fig:Exemplary-silylation-and}b),
which together with the outer silica atoms (see dark blue hexagon
in figure \ref{fig:Exemplary-silylation-and}b) are basically acting
as saturating atoms and were fixed during the geometry optimizations
in Dmol$^3$. Thus the fixed boundary of the cluster is similar to the
rigid structure of real ULK materials. 

\section{Fragmentation Reactions}

\begin{figure}[!b]
\includegraphics[width=0.45\textwidth]{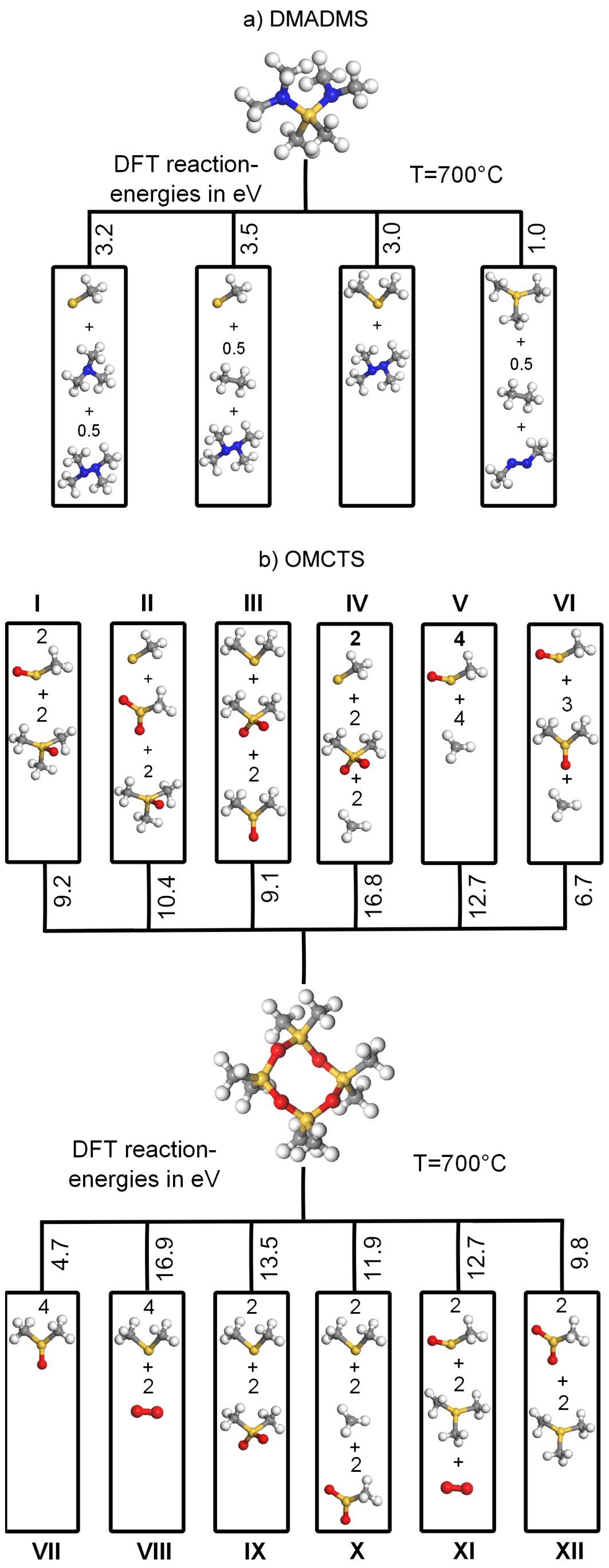}
\protect\caption{Fragmentation of silylation precursors at $T=700\deg$ C from DFT calculations.
a) Four different possible fragmentation reactions of DMADMS. b) Twelve
different possible fragmentation reactions of OMCTS. \label{fig:Fragmentation-of-DMADMS} }
\end{figure}

We focus our work on the fragmentation reactions
of the silylation precursors DMADMS and OMCTS. The DMADMS fragmentation
can only result in oxygen-free fragments. Thus, only reactions that
lead to the repair fragment $\textrm{Si}(\textrm{C}\textrm{H}_{3})_{\textrm{x}}$
and closed-shell residues were studied (figure \ref{fig:Fragmentation-of-DMADMS}a).
Here, we find that among all endothermic fragmentations, the one leading
to $\textrm{Si}(\textrm{C}\textrm{H}_{3})_{\textrm{3}}$ is the most
probable reaction path. The remaining three reactions show similar
reaction energies between 3-3.5~eV. In conclusion, for DMADMS the
chance to form small repair fragments which are prone to damages ($\textrm{Si}\textrm{C}\textrm{H}_{3}$)
is low.

We restricted the studied fragmentation reactions of OMCTS to the
ones that result to four repair fragments from figure \ref{fig:Overview-Repair-Fragments}
with possible residual molecules (figure \ref{fig:Fragmentation-of-DMADMS}b).
In accordance with the DMADMS fragmentation, the OMCTS fragmentation
reactions are also endothermic, however they require about five times
the energy of the DMADMS fragmentation reactions. This finding can
be explained easily by the fact that at least four strongly polar
SiO-bonds need to be broken in OMCTS, in contrast to only two SiN-bonds
in DMADMS.

The energetically favored reaction for OMCTS is fragmentation \textbf{VII}
into four $\textrm{Si}\textrm{O}(\textrm{C}\textrm{H}_{3})_{\textrm{2}}$
fragments as it demands the least amount of bond breaking. Among the
remaining studied fragmentations, the ones leading to at least one
$\textrm{Si}\textrm{O}(\textrm{C}\textrm{H}_{3})_{\textrm{2}}$ fragment
or fragments with three methyl groups are preferential (see reactions
\textbf{I}, \textbf{II}, \textbf{III},\textbf{ VI}, \textbf{XII}),
regardless whether they contain an additional oxygen atom. Thus, both
the DMADMS and the OMCTS fragmentation suggest that larger repair
fragments are advantageous.

Based on their dominant repair fragments, OMCTS should show a better
repair effect than DMADM\textcolor{black}{S. This is due to the fact
that $\textrm{Si}(\textrm{C}\textrm{H}_{3})_{\textrm{3}}$ as the
dominant repair fragment of DMADMS is only able to }cure exactly one
OH-damage while at the same time providing steric hindrance for further
repair reactions in its close vicinity. OMCTS's dominant repair fragment
$\textrm{Si}\textrm{O}(\textrm{C}\textrm{H}_{3})_{\textrm{2}}$, on
the other hand, is able to cure both an OH- and H-damage at the same
time, while leaving more space for other nearby repair reactions.
Thus more repair reactions should take place with OMCTS as the plasma
enhanced fragmented silylation precursor.

\begin{figure}[b]
\includegraphics[width=0.49\textwidth]{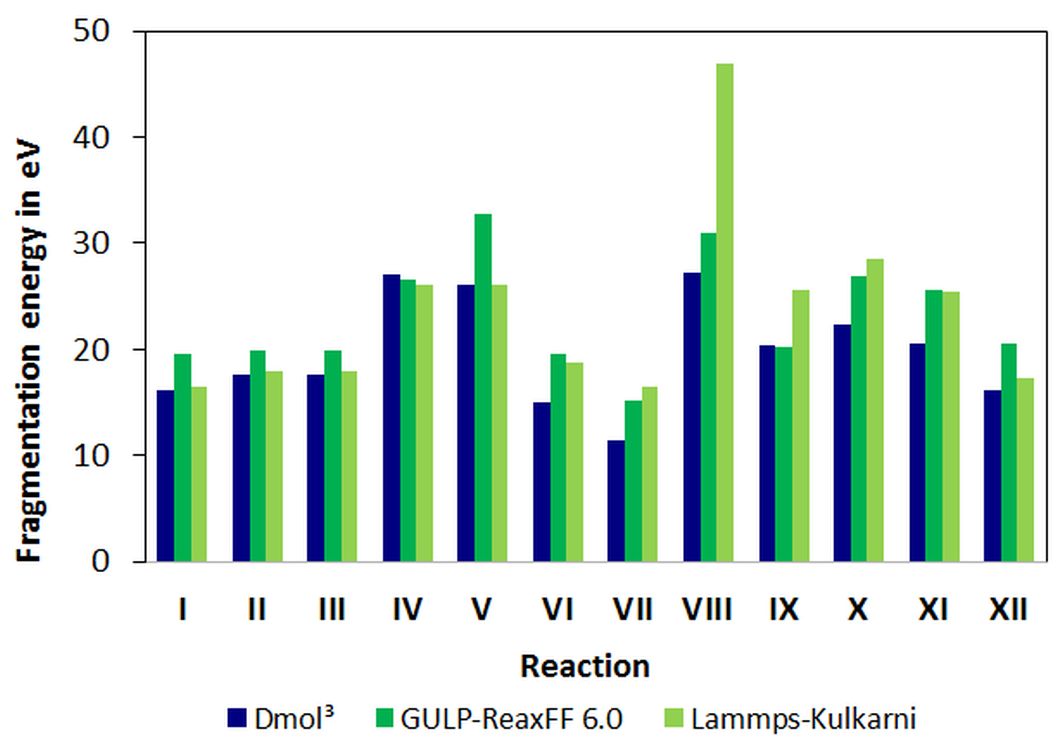}

\protect\caption{Fragmentation of OMCTS with MD and DFT. The fragmentation energies
of the twelve OMCTS fragmentations from figure \ref{fig:Fragmentation-of-DMADMS}b)
are obtained from MD and DFT calculations at T=0K.\label{fig:Frag-OMCTS-MD}}
\end{figure}

The fragmentation of OMCTS at T=0K was chosen as a test case for the
reliability of the used MD forced fields (see figure \ref{fig:Frag-OMCTS-MD}).
As to be expected, the MD results differ from the energies gained
by DFT. However, they still agree on the favored fragmentation reaction
for OMCTS (fragmentation \textbf{VII}). Further, both Gulp and Lammps
with the Kulkarni force field have an average mean squared error of
14.9 and 15.0 eV, respectively. Excluding the large derivation of
Lammps for fragmentation \textbf{VIII}, its average mean squared error
decreases to 9.4 eV. Fragmentation \textbf{VIII} is also an indicator
that the oxygen-oxygen interaction parameter in the Kulkarni force
field needs to be optimized to obtain quantitatively correct results.
In summary, it can be said that both Lammps and Gulp are nearly equally
reliable when trends in the fragmentation of silylation precursors
are to be studied. However, the use of both MD force fields is restricted
to a qualitative analysis only.

\section{Plasma repair process }

The second focus point of our study was to investigate the performance
of the repair fragments and whether MD could be used for the simulation
of the plasma repair process on more complex surface models. For example,
to study simultaneous repair processes or to analyze the diffusion
behavior of the repair fragments into a porous ULK cluster, larger
model systems with dimensions of about 10~x~10~Å would be required.
Such systems are too large to handled by DFT and thus MD with a well
parameterized force field needs to be employed. 

\begin{figure}[!th]
\includegraphics[width=0.49\textwidth]{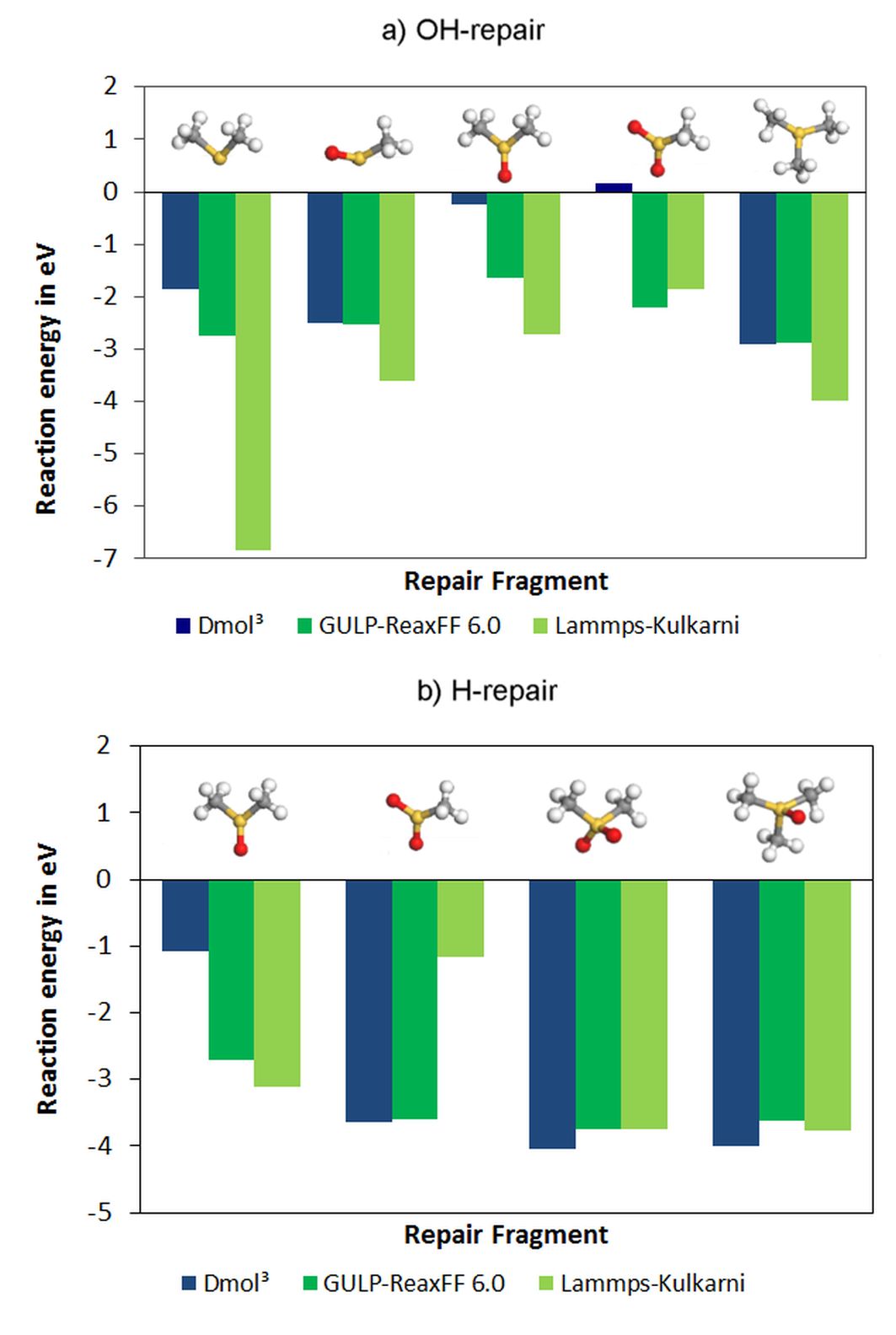}

\protect\caption{The energies of the damage repair reactions gained from Dmol$^3$, Gulp
and Lammps. a) OH-repair reactions and b) H-repair reactions. \label{fig:OH-repair-MD}}
\end{figure}

For this purpose we investigated the plasma repair process with MD
and DFT on a silicon oxide cluster as described in the section 'Model
System' and shown in figure \ref{fig:Exemplary-silylation-and}b.
We restricted our studied repair reactions to those repair fragments
that DMADMS and OMCTS favor during their fragmentations. We show the
resulting reaction energies of the repair reactions in figures \ref{fig:OH-repair-MD}a)
and b) for the OH- and H-repair, respectively.

The main conclusion that can be drawn about the reliability of the
used MD force fields is that while Lammps with the Kulkarni force
field \cite{19} provided the correct trends for the OMCTS
fragmentation, it is not suited for studying the plasma repair pro\textcolor{black}{cess,
because it unsystematically under- and overestimates the reaction
energies.}

Results from Gulp also differ from the DFT energies which are taken
as a reference. However, Gulp still conserves the general trend of
the reaction energies and thus could be use to approximately study
the plasma repair process, if a large error range of up to 2 eV is
taken into consideration. While this is sufficient for a first analysis
of the performance of the plasma repair process on the system shown
in figure \ref{fig:Exemplary-silylation-and}b, an improvement of
the existing force field is necessary to gain reliable results of
the repair behaviors or diffusion processes for more complex systems
including pores and methyl groups.

Looking at the repair reactions themselves, the results state that
(nearly) all repair reactions are highly exothermic because of the
reactive fragments produced in the plasma. Therefore, both damage
repairs will take place when the repair fragments come into contact
with the damaged ULK material. The DFT and Gulp calculations further
indicate that the repair of an H-damage is energetically more favorable
than the repair of an OH-damage.

Regarding the most effective repair fragments for the OH- and H-damage,
DFT and Gulp results are in agreement that these are $\textrm{Si}(\textrm{C}\textrm{H}_{3})_{\textrm{3}}$/$\textrm{SiO}\textrm{C}\textrm{H}_{3}$
and $\textrm{SiO}_{2}(\textrm{C}\textrm{H}_{3})_{\textrm{2}}$/$\textrm{SiO}(\textrm{C}\textrm{H}_{3})_{\textrm{3}}$/$\textrm{SiO}_{2}\textrm{C}\textrm{H}_{3}$,
respectively. Among these repair fragments, $\textrm{SiO}\textrm{C}\textrm{H}_{3}$
and \textcolor{black}{$\textrm{SiO}_{2}\textrm{C}\textrm{H}_{3}$
have the disadvantage that they are prone to develop new defects (refer
to figure \ref{fig:Overview-Repair-Fragments}). The remaining repair
fragments, on the other hand, are rather large and thus will prevent
other repair processes to take place in the close vicinity. This means
that while the repair reactions are all strongly exothermic, not a
100\% repair of the damage will be achieved due to steric hindrances
and the susceptibility of the repair fragments with only one methyl
group to form new defects.}

If the above listed repair fragments are compared to the dominant
fragments of the OMCTS and DMADMS fragmentation, then it can be seen
that \textcolor{green}{{} }\textcolor{black}{$\textrm{SiO}(\textrm{C}\textrm{H}_{3})_{\textrm{2}}$
as the dominant repair fragment of OMCTS is o}ne of the least effective
repair fragments, \textcolor{black}{while $\textrm{Si}(\textrm{C}\textrm{H}_{3})_{\textrm{3}}$
as the dominant repair fragment of DMADMS is t}he most effective OH-damage
repair fragment. However, the energetically next favored fragments
of OMCTS (refer to reactions \textbf{I}, \textbf{II}, \textbf{VI}
and \textbf{XII} in figure \ref{fig:Fragmentation-of-DMADMS}b) are
all very effective plasma repair fragments for both the OH- and H-damage.
Thus, OMCTS should still demonstrate a good repair performance, which
is even better than the repair effect of DMADMS. The last assumption
can be made because the reaction energy for the H-repair is over 1~eV
higher than for the OH-repair, which is the only damage DMADMS can
cure.

\subsection{\textcolor{black}{Comparison with ULK-fragments}}

\begin{figure}[!th]
\includegraphics[width=0.49\textwidth]{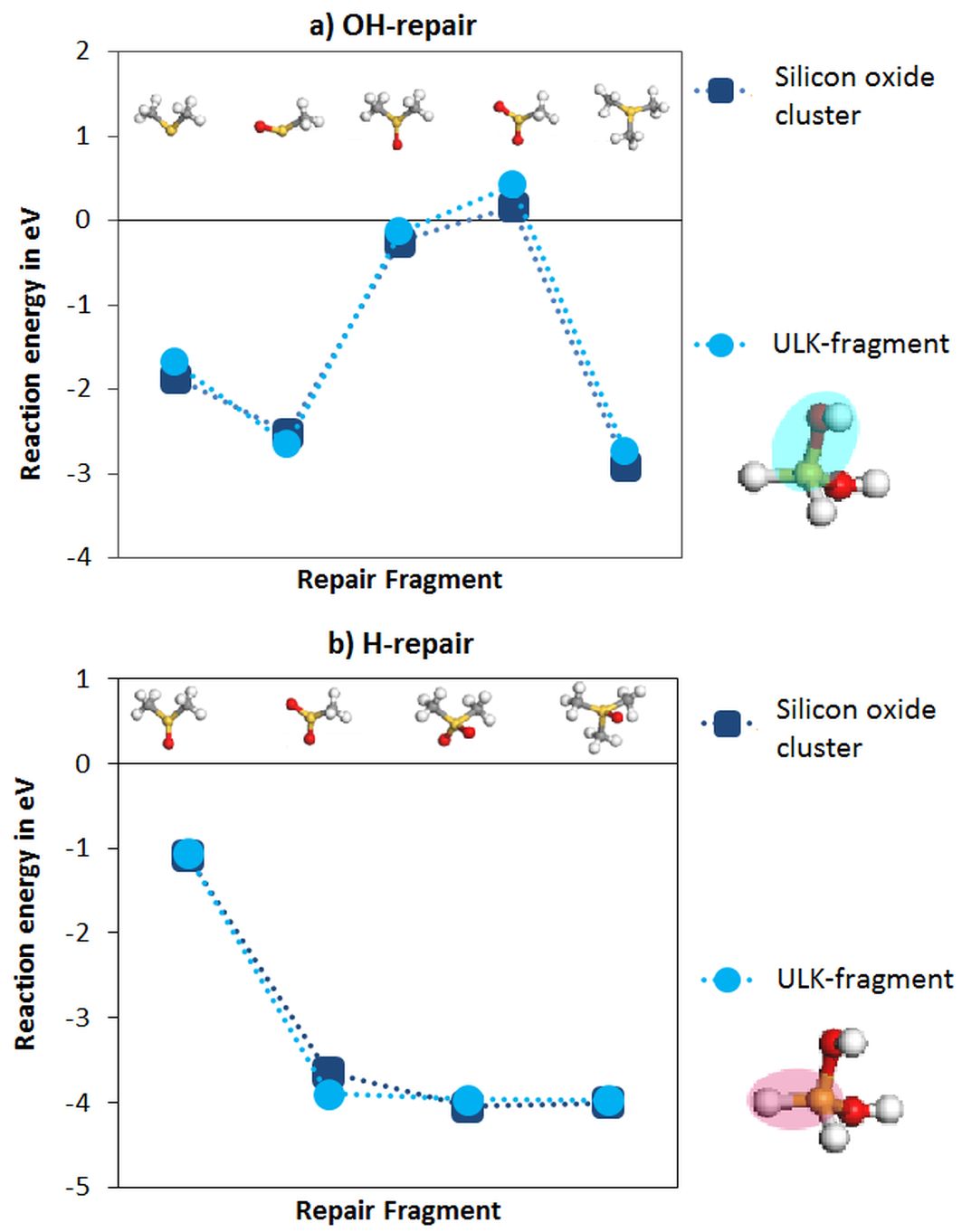}

\protect\caption{\textcolor{black}{Comparison of the results of the silicon oxide cluster
and the ULK-fragment for a) the repair of the OH-damage and b) the
repair of the H-damage. The a) blue and b) pink highlighted area of
the ULK-fragment shows the OH- and H-damage, respectively. The dotted
lines are only a guide for the eyes and show no correlations.}\textcolor{green}{{}
\label{fig:ULK-fragment}}}
\end{figure}

\textcolor{black}{In \cite{20} we studied the repair
performance of all repair fragments shown in figure \ref{fig:Overview-Repair-Fragments}
with identical DFT settings but with another, much smaller model system.
This model system consists of small ULK-fragments (see example in
figure \ref{fig:ULK-fragment}), which possess only one silicon atom
together with OH- and H-defects as well as saturating H atoms. In
analogy to the silicon oxide cluster, a part of the ULK-fragment was
fixed during geometry optimizations. Specifically, all atoms outside
the in blue and pink highlighted areas in figure \ref{fig:ULK-fragment}
were fixed. }

\textcolor{black}{In figure \ref{fig:ULK-fragment} we compare the
repair reactions of this study with the results from \cite{20}.
The motivation for this comparison is the unsatisfactory performance
of the MD force fields and the high numerical effort of the DFT calculations
for the silicon oxide cluster. For both OH- and H-repair reactions
the results with the small ULK-fragment and with the extended cluster
model agree very well energetically. Thus, when assessing only the
effectiveness of new repair fragments the computationally less demanding
ULK-fragments can be employed. The ULK-fragment model system is sufficient
to quickly and at the same time qualitatively correctly describe the
repair behavior of plasma fragmented silylation precursors.}

\subsection{Comparison with Experiments}

In a final step we compare our theoretical results with the experiments
carried out by K{\"o}hler et al. \cite{08}. They
investigated the repair with plasma enhanced fragmented OMCTS and
DMADMS, in particular the repair of surface and sidewall damages.
K{\"o}hler et al. found that both fragmented silylation precursors minimize
the surface damage, with OMCTS showing an improved repair performance
compared to DMADMS. This is in agreement with our result that the
repair of H-damages and thus oxygen-containing repair fragments possess
more strongly exothermic reaction energies and thus the use of OMCTS
results in a better repair effect.

With the repair of sidewall damages K{\"o}hler et al. investigated the
correlation between the repair process and OMCTS-flow together with
the use of the gas additives oxygen or methane. The presence of oxygen
led to a highly improved repair of the damages. This can be understood
by our result that only oxygen containing repair fragments yield exothermic
reaction energies for the repair of H-damages. Thus, the repair of
H-damages is strongly increased by the availability of oxygen.

\section{Conclusion}

We have shown that the fragmentation of DMADMS requires less energy
than the OMCTS fragmentation. However, their dominant fragmentation
reactions and the resulting primary repair fragments, indicate that
the plasma repair will be less effective for DMADMS in comparison
to OMCTS. Further, (nearly all of) the favored fragments from the
OMCTS fragmentation are very effective repair fragments for both the
OH- and H-damage, while the favored repair fragment of DMADMS can
only repair OH-damages. These results are in good agreement with the
experimental study carried out by K{\"o}hler et al.

By comparing DFT and MD results we can conclude that for using MD
to simulate the plasma repair process with larger and more realistic
silicon oxide structures which include pores, the existing force fields
need to be optimized. However, Gulp with the ReaxFF 6.0 force field
can be used for a first investigation of the qualitative trends.
\begin{acknowledgments}
S. Gemming acknowledges funding from the Initiative and Networking
Funds of the President of the Helmholtz Association via the W3 program.
C. Wagner and S. Gemming acknowledge the funding by the DFG research
unit 1713 \textquotedbl{}Sensoric Micro- and Nanosystems\textquotedbl{}.
\end{acknowledgments}

\bibliographystyle{elsarticle-num}

\end{document}